\documentclass[useAMS,usenatbib]{mn2e}
\usepackage{graphicx}
\usepackage{latexsym}
\usepackage{psfig}
\usepackage{amssymb,amsmath}
\usepackage{subfigure}

\title[The spatial distributions of stars and gas]{On the spatial distributions of stars and gas in numerical simulations of molecular clouds}

\author[R. J. Parker \& J. E. Dale]{Richard  J. Parker$^{1}$\thanks{E-mail: R.J.Parker@ljmu.ac.uk} and James E. Dale$^{2,3}$ \vspace*{0.1cm}\\
$^{1}$Astrophysics Research Institute, Liverpool John Moores University, 146 Brownlow Hill, Liverpool, L3 5RF, UK\\
$^{2}$Excellence Cluster `Universe', Boltzmannstra{\ss}e 2, 85748 Garching, Germany\\
$^{3}$Universit{\"a}ts-Sternwarte M{\"u}nchen, Scheinerstra{\ss}e 1, 81679 M{\"u}nchen, Germany
}

\begin{document}

\pagerange{\pageref{firstpage}--\pageref{lastpage}} \pubyear{2015}

\date{}

\maketitle

\label{firstpage}

\def\mnras{MNRAS}
\def\apj{ApJ}
\def\aj{AJ}
\def\aap{A\&A}
\def\apjl{ApJL}
\def\apjs{ApJS}
\def\araa{ARA\&A}
\def\pasj{PASJ}
 
\begin{abstract}
We compare the spatial distribution of stars which form in hydrodynamical simulations to the spatial distribution of the gas, using the $\mathcal{Q}$-parameter. The $\mathcal{Q}$-parameter enables a self-consistent comparison between the stars and gas because it uses a pixelated image of the gas as a distribution of points, in the same way that the stars  (sink particles in the simulations) are a distribution of points. We find that, whereas the stars have a substructured, or hierarchical spatial distribution ($\mathcal{Q} \sim 0.4 - 0.7$), the gas is dominated by a  smooth, concentrated component and typically has $\mathcal{Q} \sim 0.9$. We also find no statistical difference between the structure of the gas in simulations that form with feedback, and those that form without, despite these two processes producing visually different distributions. These results suggest that the link between the spatial distributions of gas, and the stars which form from them, is non-trivial. 
\end{abstract}


\section{Introduction}

The physics of star formation results in stars grouped together in regions which exceed the mean density of the Galactic disc by several orders of magnitude \citep{Blaauw64,Lada03,Porras03,Bressert10}. Furthermore, there is growing evidence that star formation is hierarchical from the interstellar medium (ISM) down to sub-pc scales \citep{Hoyle53,Scalo85,Efremov95,Elmegreen06b,Bastian07,Kruijssen12b}, and that there is no preferred spatial scale for a given star formation `event'. 

Understanding this complex hierarchical picture of star formation requires analysis of the spatial distribution of gas from the ISM down to the giant molecular clouds (GMCs) from which stars form, to the substructure of the clouds and then of the spatial distribution of the stars themselves. In particular, do stars exhibit the same spatial distribution as the ISM \citep[e.g. as argued by][for the NGC346 star-forming region]{Gouliermis14} and if so, does this provide information on the physical processes from the ISM down to individual star formation? Answering this question requires the study of star-forming regions at the earliest possible stages, since the gas structure is likely to be disrupted by feedback and the stellar structure by dynamical interactions. Either or both of these processes can in principle erase any spatial correlation between gas and stars on short timescales.

In order to address these questions, analysis of both gas in the ISM and GMCs, and stars in young regions, must be undertaken using a self-consistent method. In recent years, structural analysis of stars in young regions has been performed using the powerful $\mathcal{Q}$-parameter \citep[e.g.][]{Cartwright04,Bastian09,Schmeja08,Sanchez09,Hetem15}, which combines information on the minimum spanning tree of a distribution with the typical separation between the points in the distribution. This technique has also been developed to study the gas distribution in images, by appropriately weighting the flux from pixels in an image to create a distribution of points \citep*{Lomax11}.

{It is not clear to what extent and for how long the spatial distribution of stars should follow the same distribution as the gas from which they form, and this has yet to be addressed observationally.} However, detailed hydrodynamical simulations of star formation provide both information on the spatial distribution of gas, and on the distribution of stars (sink particles). In this paper, we examine the hydrodynamical simulations of star formation by \citet{Dale14} and measure the spatial distribution of both the gas (from pixelated images) and the distribution of the sink particles, using the $\mathcal{Q}$-parameter. The suite of simulations by \citeauthor{Dale14} include control runs without feedback from photoionisation and stellar winds, and so we also search for differences in the gas distributions for clouds that are influenced by feedback mechanisms, and those that are not.  

The paper is organised as follows. In Section~\ref{methods} we decribe the implementation of the $\mathcal{Q}$-parameter on gas distributions, in Section~\ref{results} we present our results, in Section~\ref{discuss} we provide a discussion and we conclude in Section~\ref{conclude}.

\section{Method}
\label{methods}

The simulations of \citeauthor{Dale14} use a Smoothed--Particle Hydrodynamics (SPH) code to model the evolution of GMCs of a range of sizes and masses and are seeded with Burgers turbulence such that their initial virial ratios are either 0.7 or 2.3. The formation of stars is followed using sink particles \citep{Bate95} which, in the simulations analysed here, have accretion radii of 0.005\,pc and can be regarded approximately as individual stars.

\citeauthor{Dale14} allow their simulations to evolve until three objects exceed 20\,M$_{\odot}$ at which point each calculation is forked into a control run and a feedback run. The control runs evolve purely hydrodynamically as before, while the feedback runs are impacted by the ionising radiation and/or stellar winds of the massive stars, modelled respectively by the algorithms presented in \cite{Dale07,Dale12a} and \cite{Dale08}. All runs are continued for as near as possible to 3\,Myr to evaluate the effects of pre-supernova feedback on the clouds and clusters.

For the analysis presented here, pixelated column-density maps are constructed by drawing a pixel grid over the simulation, placing each SPH particle on the grid and column-integrating through its smoothing kernel onto all grid cells whose centres it overlaps. Since the maps presented here have a lower resolution than the local resolution of the hydrodynamic simulations, particles which are too small to overlap the centre of any grid cells have their mass smeared out over the area of the cell in which they lie.

We use the $\mathcal{Q}$-parameter \citep{Cartwright04,Cartwright09b} to quantify the spatial distribution of both stars and gas in the hydrodynamical simulations of \citet{Dale14}. The $\mathcal{Q}$-parameter is determined by constructing a minimum spanning tree (MST) of all of the points in a distribution and then dividing the mean MST branch length, $\bar{m}$ by the mean separation between points, $\bar{s}$:
\begin{equation}
\mathcal{Q} = \frac{\bar{m}}{\bar{s}}.
\end{equation}  
Determining $\mathcal{Q}$ for the stars (i.e.\,\,sink particles) in the simulations is trivial, but determining $\mathcal{Q}$ for the gas is more involved \citep*{Cartwright06}. \citet*{Lomax11} provide a method for converting the flux distribution in a pixelated image into a distribution of points, from which the $\mathcal{Q}$-parameter can then be calculated. We refer the interested reader to \citet{Lomax11} for full details of the method, including examples of its use on synthetic images, but briefly summarise the method here. In the simulations of  \citet{Dale14}, we use gas column density as the flux.

\begin{figure}
\begin{center}
\rotatebox{270}{\includegraphics[scale=0.4]{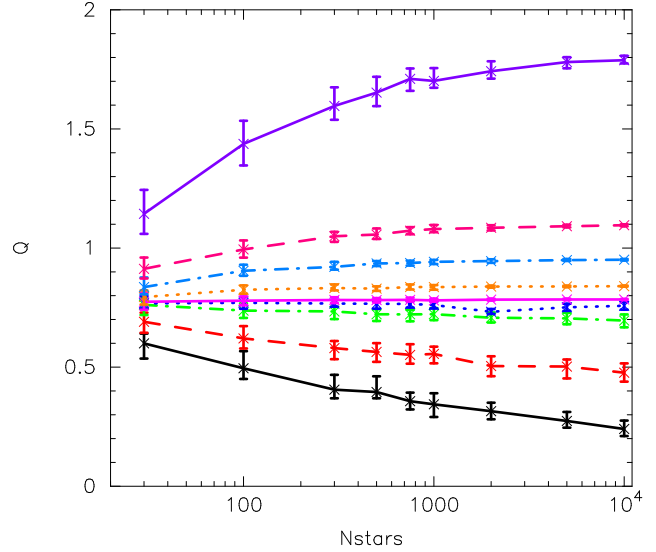}}
\end{center}
\caption[bf]{Dependance of the $\mathcal{Q}$-parameter on the number of particles in a distribution. From top to middle, the lines indicate a  smooth, concentrated distribution with power law density profile $n \propto r^{-2.9}$ (solid purple),  $r^{-2.5}$ (dashed raspberry), $r^{-2.0}$ (dot-dashed pale blue), $r^{-1.0}$ (dotted orange), $r^{0}$ (solid magenta); and from bottom to middle the lines indicate a substructured fractal distribution with fractal dimension $D = 1.6$ (solid black), 2.0 (dashed red), 2.6 (dot-dashed green), 3.0 (dotted dark blue). The error bars indicate the interquartile range from 100 identical realisations of the same cluster for a given number of points.}
\label{Qpar_all}
\end{figure}

For an $N_{\rm pix} = \mathcal{I}  \times \mathcal{J}$ array of pixels, the total flux received from the pixels, $F_{\rm tot}$ is:
\begin{equation}
F_{\rm tot} = \sum\limits_{i=1}^{i=\mathcal{I}} \sum\limits_{j=1}^{j=\mathcal{J}} F_{ij}.
\end{equation} 
In order to convert the flux distribution into a distribution of points, \citet{Lomax11} then define the flux quantum as 
\begin{equation}
\Delta F = \frac{F_{\rm tot}}{N_{\rm pix}}.
\end{equation}
We start by choosing a pixel, $\mathcal{R}_{ij}$ at random. If $F_{ij} \geq \Delta F$, then the flux in that pixel is reduced
\begin{equation}
F_{ij} \rightarrow F_{ij} - \Delta F
\end{equation}
and we place a point at ${\bf r}_{ij} + \Delta {\bf r}_{\rm rnd}$, where ${\bf r}_{ij}$ is the centre of pixel $\mathcal{R}_{ij}$ and $ \Delta {\bf r}_{\rm rnd}$ is a small random displacement an order of magnitude smaller than the pixel size, to prevent the final point distribution from having a gridded appearance.

If $F_{ij} < \Delta F$, then we consider a patch of $n$ pixels \citep[see fig.\,1 in][]{Lomax11}. $n$ is increased until the flux from the patch is equal to, or exceeds $\Delta F$:
\begin{equation}
F_{n-{\rm patch}} = \sum\limits_{n-{\rm patch}} F_{ij} \geq \Delta F.
\end{equation} 
The flux from each pixel is then uniformly reduced
\begin{equation}
F_{ij} \rightarrow F_{ij}\left(1 - \frac{\Delta F}{F_{n-{\rm patch}}}\right),
\end{equation}
and a point is placed at position ${\bf r}_{\rm pnt}$ which is equal to the weighted centre of the removed flux, plus a small random displacement:
\begin{equation}
{\bf r}_{\rm pnt} = \sum\limits_{n-{\rm patch}} \left\{\frac{F_{ij} {\bf r}_{ij}}{F_{n-{\rm patch}}}\right\} + \Delta {\bf r}_{\rm rnd}.
\end{equation} 
This process is repeated $N_{\rm pix}$ times, whereafter the total flux is now zero. 

\begin{figure*}
  \begin{center}
\setlength{\subfigcapskip}{10pt}
%\hspace*{-1.cm}
\subfigure[Raw column density image of the gas and sinks (white dots) in the control Run I simulation. For the sink particles, $\mathcal{Q} = 0.72$.]{\label{control_run-a}\rotatebox{0}{\includegraphics[scale=0.63]{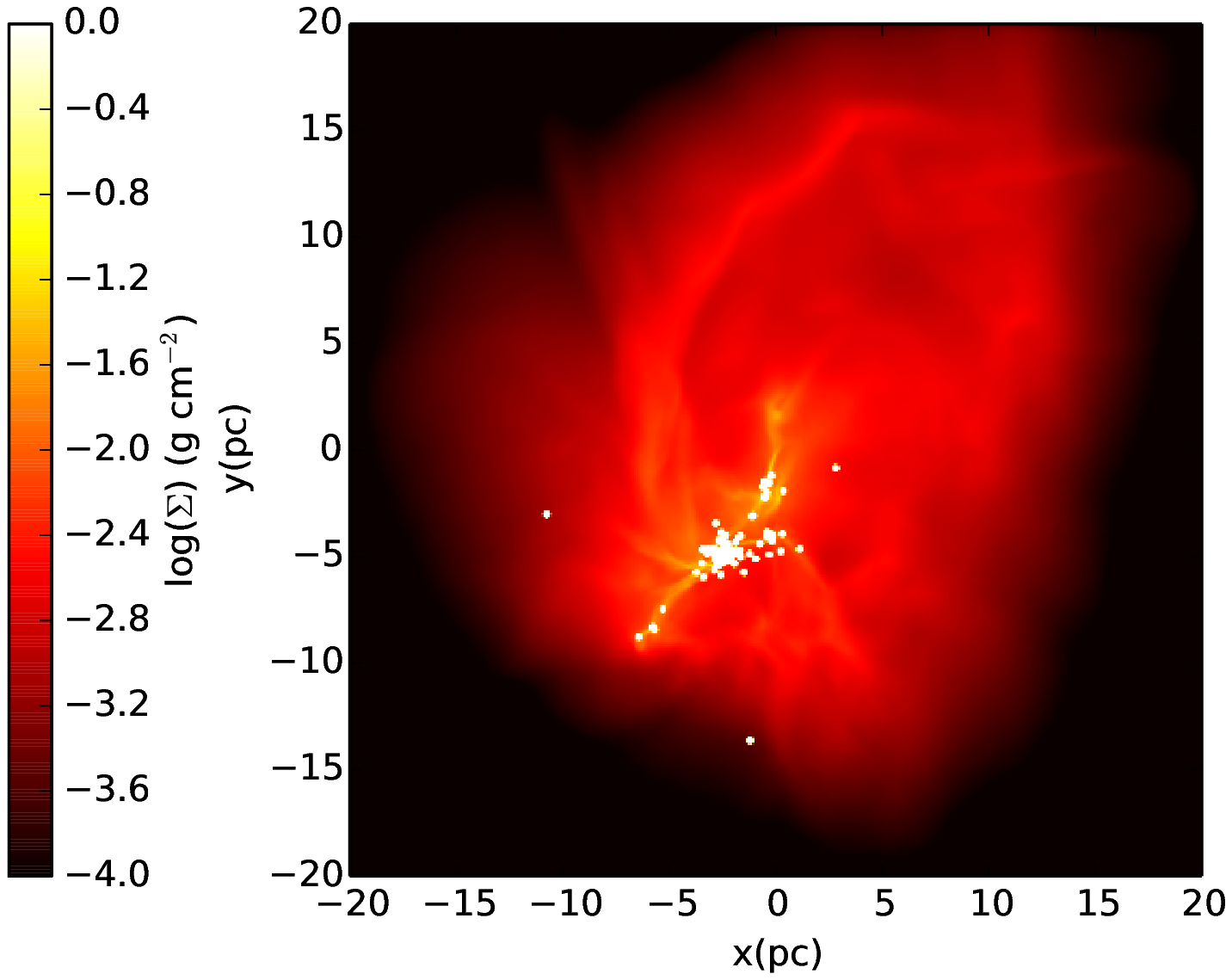}}}
%\vspace*{-2cm} 
%\hspace*{0.3cm}
\hspace*{-0.0cm}
\subfigure[Particle point representation of the gas only in the control Run I simulation at a resolution of 128 x 128 pixels; $\mathcal{Q} = 1.01$.]{\label{control_run-b}\rotatebox{0}{\includegraphics[scale=0.375]{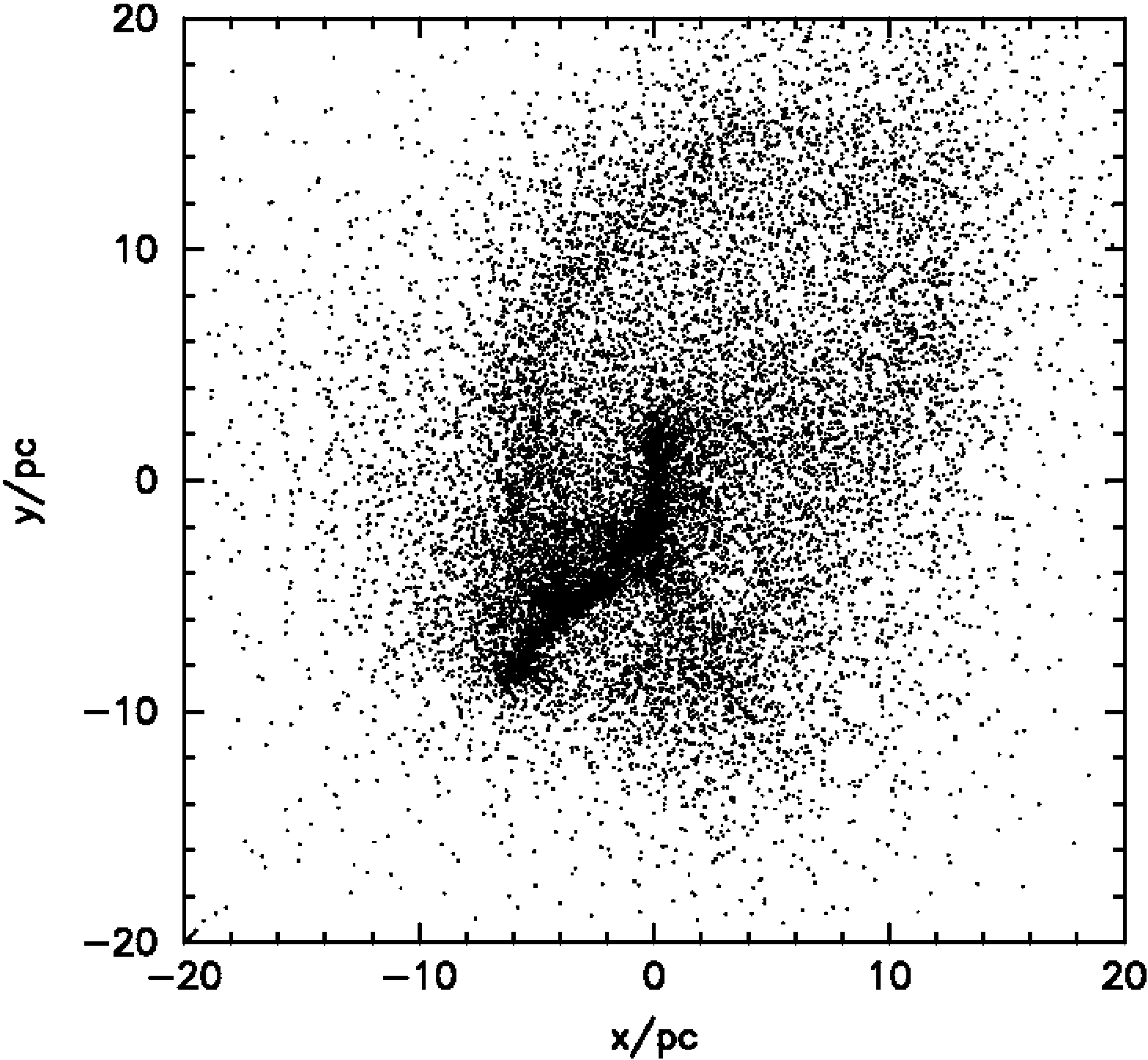}}}
\caption[bf]{The spatial distribution of stars and gas in the control Run I simulation from \citet{Dale14} 2.2 Myr after the point where the calculation is forked into separate dual-feedback and control runs. In panel (a) we show the surface density distribution the gas, and the sink particles are shown by the white points. In panel (b) we show the pixel point distribution for the gas. The sink particles have $\mathcal{Q} = 0.72$ and the gas distribution has $\mathcal{Q} = 1.01$.}
\label{control_run}
  \end{center}
\end{figure*}

We now have a distribution of  $N_{\rm pix}$ points with which we calculate the $\mathcal{Q}$-parameter for the gas distribution. The $\mathcal{Q}$-parameter provides a measure of the degree to which a distribution is substructured, or  concentrated. \citet{Cartwright04}, \citet{Cartwright09b} and \citet{Lomax11} provide calibration data for synthetic models, so for a given  $\mathcal{Q}$-parameter one can assign a fractal dimension (in the substructured case), or the radial density profile exponent (in the  smooth, concentrated case). 

However, as mentioned by \citet{Lomax11}, the $\mathcal{Q}$-parameter varies slightly depending on the number of points in the distribution. This is usually not a problem when comparing e.g.\,\,the outcome of simulations to real star-forming regions, as the numbers of stars are similar (between 100 -- 1000). However, when constructing a point distribution from pixels, in order to properly sample the gas distribution we find at least $64 \times 64 = 4096$ pixels are required, but $128 \times 128 = 16384$ pixels are desirable. 

In Fig.~\ref{Qpar_all} we show the dependence of  $\mathcal{Q}$ on the number of points, $N$. The coloured lines from top to middle indicate a  smooth, concentrated distribution with power law density profile  $n \propto r^{-2.9}$ (solid purple),  $r^{-2.5}$ (dashed raspberry), $r^{-2.0}$ (dot-dashed pale blue), $r^{-1.0}$ (dotted orange), $r^{0}$ (solid magenta). From bottom to middle, the lines indicate a substructured fractal distribution with fractal dimension $D = 1.6$  (solid black), 2.0 (dashed red), 2.6 (dot-dashed green), 3.0 (dotted dark blue). The error bars indicate the interquartile range from 100 statistically identical realisations of the same cluster for a given number of points. $\mathcal{Q}$ tends to more extreme values with increasing $N$ if a distribution is highly substructured, or highly  concentrated, but remains roughly constant for more moderate distributions. In order to check that these results are not simply due to sampling too few points, we created each dataset with $10^4$ points and then randomly chose $N$ from this distribution and found  very similar results.

\section{Results}
\label{results}

We analyse five sets of SPH simulations from \citet{Dale14}. These simulations follow the evolution of giant molecular clouds as they form stars without feedback from photoionisation and stellar winds (`control runs') and simulations in which feedback is switched on (`dual feedback runs'). In earlier work \citep*{Parker13a,Parker15a}, we showed that the presence (or not) of feedback can effect the long term spatial evolution of a star-forming region and here we investigate whether the spatial distribution of the gas is influenced by feedback, and if the spatial distribution of the gas follows that of the stars. 

\begin{figure*}
  \begin{center}
\setlength{\subfigcapskip}{10pt}
%\hspace*{-0.3cm}
\subfigure[Raw column density image of the gas and sinks (white dots) in the dual--feedback Run I simulation. For the sink particles, $\mathcal{Q} = 0.49$.]{\label{dual_run-a}\rotatebox{0}{\includegraphics[scale=0.63]{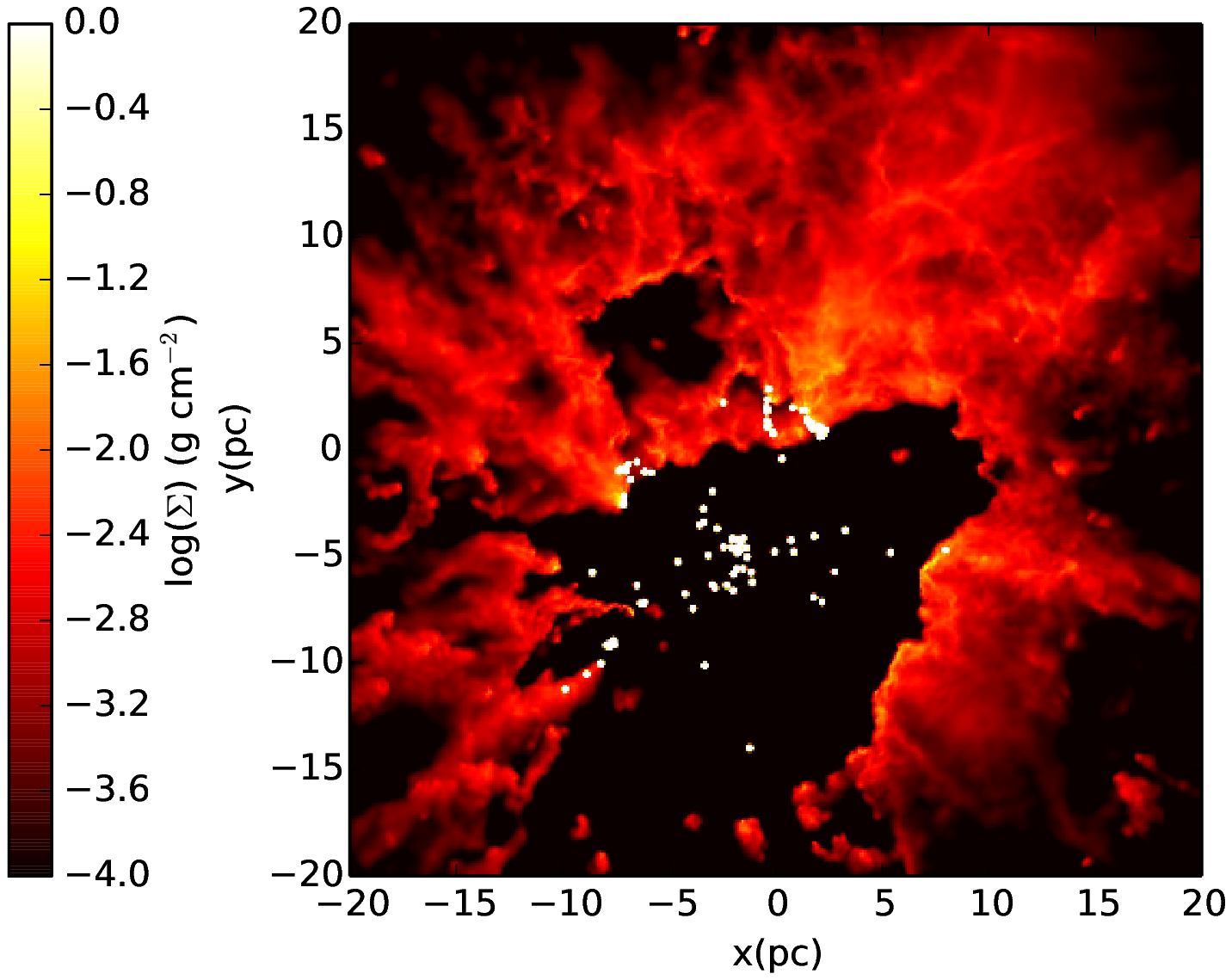}}} 
\hspace*{-0.0cm}
\subfigure[Particle point representation of the gas only in the dual--feedback Run I simulation at a resolution of 128 x 128 pixels; $\mathcal{Q} = 0.88$.]{\label{dual_run-b}\rotatebox{0}{\includegraphics[scale=0.375]{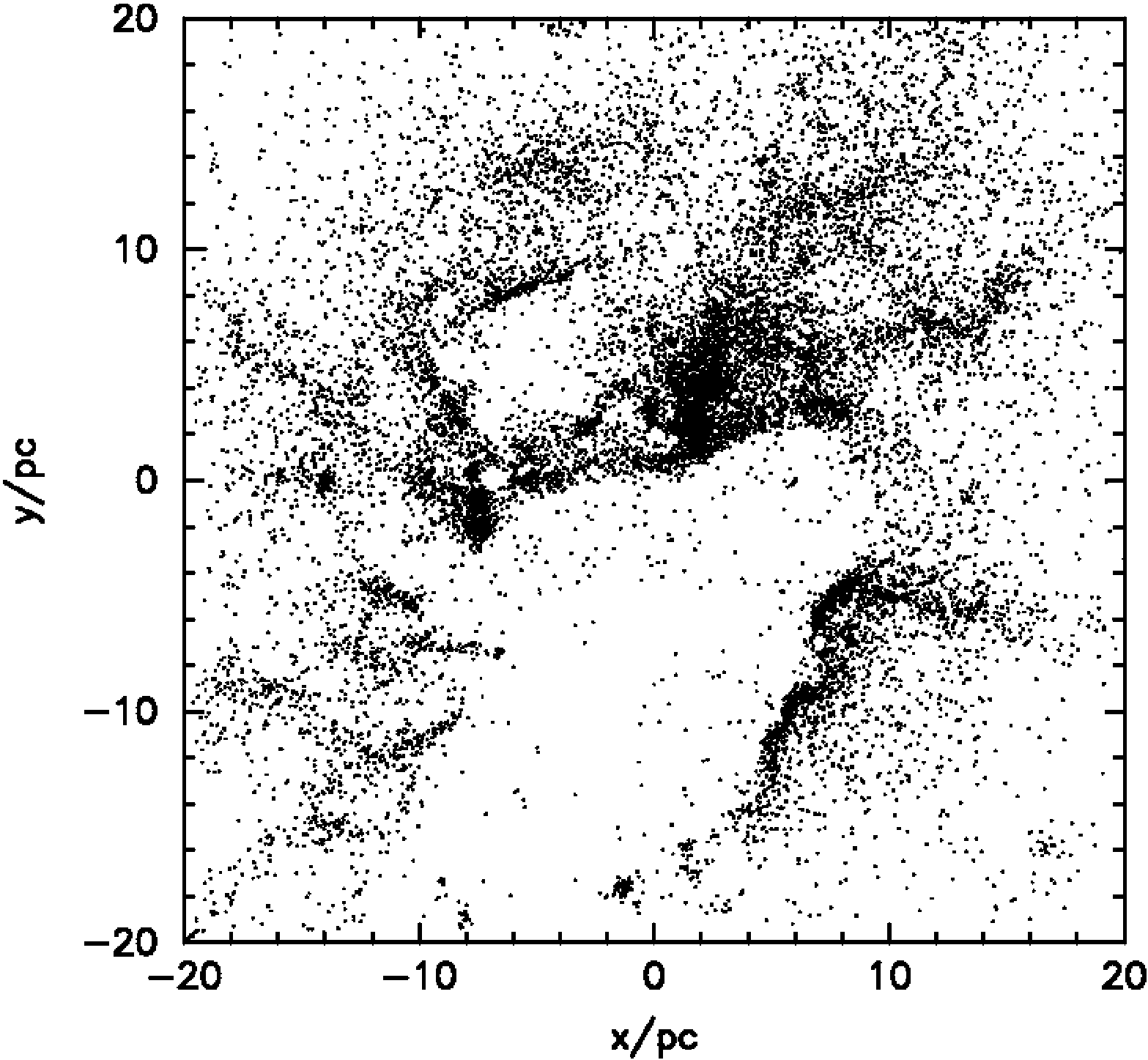}}}
\caption[bf]{The spatial distribution of stars and gas in the dual-feedback Run I from \citet{Dale14}, 2.2\,Myr after the enabling of photoionisation and winds from the O-type stars. In panel (a) we show the surface density distribution of the gas, and the sink particles are shown by the white points. In panel (b) we show the pixel point distribution for the gas. The sink particles have $\mathcal{Q} = 0.49$ and the gas distribution has $\mathcal{Q} = 0.88$.}
\label{dual_run}

  \end{center}
\end{figure*}

\begin{table*}
\caption[bf]{Measured $\mathcal{Q}$-parameters for the gas distributions in the simulations from \citet{Dale14}. The columns display the run ID, whether feedback was switched on, the $\mathcal{Q}$-parameter for the gas distribution using 64 x 64 pixels,  the $\mathcal{Q}$-parameter for the gas distribution using 128 x 128 pixels, the $\mathcal{Q}$-parameter using 128 x 128 pixels for cold gas only, the $\mathcal{Q}$-parameter using 128 x 128 pixels for cold gas with a column density above $1 \times 10^{-4}$\,g\,cm$^{-2}$, the $\mathcal{Q}$-parameter for the sink particles, and the number of sinks.}
\begin{center}
\begin{tabular}{|c|c|c|c|c|c|c|c|}
\hline 
Run ID & Feedback & $\mathcal{Q}_{\rm gas, 64~x~64}$ & $\mathcal{Q}_{\rm gas, 128~x~128}$ & $\mathcal{Q}_{\rm cold~gas, 128~x~128}$ & $\mathcal{Q}_{\rm cold~gas,~flux~limited, 128~x~128}$ & $\mathcal{Q}_{\rm sinks}$ & $N_{\rm sinks}$\\
\hline
 I & Off & 1.03 & 1.01 & 1.01  & 0.96  & 0.72 & 186\\
 I & Dual & 0.87 & 0.88 & 0.83 & 0.81 & 0.49 & 132\\
 I & Before & 1.02 & 0.99 & 0.99  & 0.95 & 0.42 & 44\\
\hline
 J & Off & 0.91 & 0.89 & 0.89 & 0.89 & 0.49 & 578 \\
 J & Dual & 0.87 & 0.90 & 0.91 & 0.92 & 0.70 &564 \\
\hline
 UF & Off & 0.88 & 0.86 & 0.86 & 0.84 & 0.77 & 66\\ 
 UF & Dual & 0.83 & 0.85 & 0.88 & 0.89 & 0.49 & 93\\
\hline
 UP & Off & 0.92 & 0.88 & 0.88 & 0.86 & 0.49 & 340\\
 UP & Dual & 0.87 & 0.88 & 0.92 & 0.88 & 0.64 & 343\\
\hline
 UQ & Off & 0.91 & 0.86 & 0.86 & 0.88 & 0.70 & 48\\ 
 UQ & Dual & 1.07 & 0.83 & 0.78 & 0.77 & 0.45 & 77\\
\hline
\end{tabular}
\end{center}
\label{cluster_props}
\end{table*}

The results from the control Run I simulation 2.2 Myr after the epoch when the calculation is split into parallel control and dual-feedback incarnations are shown in Fig.~\ref{control_run}. In Fig.~\ref{control_run-a} we show the surface density distribution of the gas.  We also show the positions of the sink particles (the white points), which have $\mathcal{Q} = 0.72$, suggesting a slightly substructured distribution. In Fig.~\ref{control_run-b} we show the distribution of points from the pixel distribution of column density, determined using the method from \citet{Lomax11}. This gas distribution has   $\mathcal{Q} = 1.01$, suggesting a  smooth, concentrated distribution.

We show the results at the same timestep of the dual-feedback Run I calculation from \citet{Dale14} which includes the effects of photoionisation feedback and stellar winds in Fig.~\ref{dual_run}. The sink particles (shown by the white points in  Fig.~\ref{dual_run-a}) have $\mathcal{Q} = 0.49$, suggesting a substructured distribution. From inspection of  Fig.~\ref{dual_run-b}, the distribution of gas also appears substructured; however, the $\mathcal{Q}$-parameter is 0.88, indicating a  smooth, concentrated distribution. 

The results for all five sets of simulations are summarised in Table~\ref{cluster_props} (all five sets have very similar morphologies). We present the $\mathcal{Q}$-parameter for a pixelated map of the gas distribution for each simulation using $64 \times 64$ pixels, and $128 \times 128 $ pixels (differences between the two are minimal). In the $128 \times 128 $ pixel case, we also present the $\mathcal{Q}$-parameter for cold gas only (i.e.\,\,ionised particles have been removed), and for cold gas above a column density threshold of  $> 1 \times 10^{-4}$\,g\,cm$^{-2}$ (this process sets the column density of around 25\,per cent of the pixels to zero). Finally, we give the $\mathcal{Q}$-parameter for the sink particles, and the number of sink particles in each simulation.

\begin{figure}
\begin{center}
\rotatebox{270}{\includegraphics[scale=0.4]{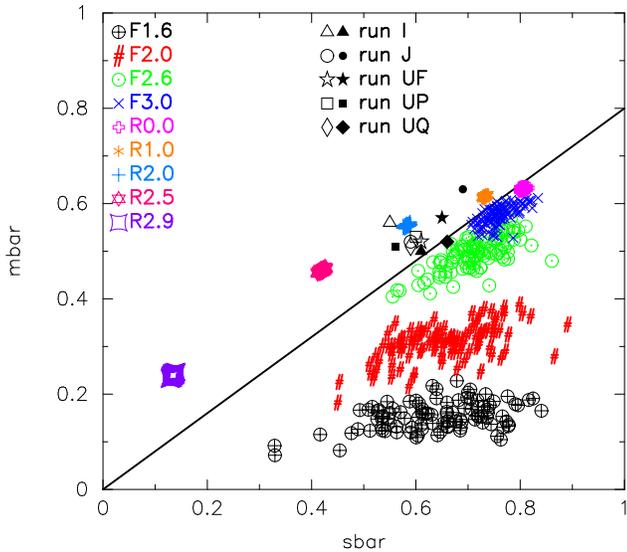}}
\end{center}
\caption[bf]{The $\bar{m} - \bar{s}$ plot \citep{Cartwright09b} for different morphologies containing $N = 10000$ points. We create 100 realisations of each morphology. Anti-clockwise from the bottom: fractals with fractal dimension $D = 1.6$ (the black crossed circles), $D = 2.0$ (the red hashtags), $D = 2.6$ (the green dotted circles), $D = 3.0$ (the dark blue crosses); then radially smooth clusters with power law density profiles $n \propto r^{0}$ (the magenta plus symbols),  $n \propto r^{-1.0}$ (orange asterisks), $n \propto r^{-2.0}$ (blue plus symbols),  $n \propto r^{-2.5}$ (raspberry hexagrams) and $n \propto r^{-2.9}$ (purple compressed squares). We overplot the  $\bar{m}$ and $\bar{s}$ values for the pixel-point distributions for the gas in the simulations. Control runs are shown by the open symbols, dual feedback runs are shown by the filled symbols. Finally, values for $\bar{m}$ and $\bar{s}$ corresponding to a $\mathcal{Q}$-parameter of 0.8 are shown by the solid black line.  }
\label{cartwright_plot}
\end{figure}

For one simulation set (Run I) we also present the $\mathcal{Q}$-parameters for the simulations at the point before feedback is switched on. The gas distribution in this simulation is (as we might expect) similar to the end of the control run ($\mathcal{Q}_{\rm gas} = 1.02$), but the $\mathcal{Q}$-parameter for the sink particles is similar to the value at the end of the dual-feedback run $\mathcal{Q}_{\rm sinks} = 0.42$. This is likely due to the simulated cluster being substructured early on, and the control run erasing some of that substructure due to dynamical evolution \citep[as discussed in][]{Parker13a}.

Two main results are apparent in our analysis. Firstly, the $\mathcal{Q}$-parameter for the sink particles is systematically lower than the $\mathcal{Q}$-parameter of the gas distribution. [Note that this is not due to the differences in the numbers of particles used to determine the $\mathcal{Q}$-parameter, as highlighted in Fig.~\ref{Qpar_all}; this effect is only important if both $\mathcal{Q}$-values are in the same regime (i.e. substructured, or  concentrated)]. Typically, $\mathcal{Q}_{\rm gas} \sim 0.9$ (indicating a  smooth, concentrated distribution), whereas the sink particles tend to have $\mathcal{Q}$-parameters between 0.4 -- 0.7 (indicating a substructured, or hierarchical distribution).

Secondly, and perhaps most strikingly, differences between the gas distributions in the simulations with and without feedback are minimal in most cases (despite the apparently substructured gas distribution for the runs in which feedback is switched on). If we compare the value of $\mathcal{Q}_{\rm gas}$ for the control run simulations, the range of values is 0.86 -- 1.01 for the 128 x 128 pixel maps. The dual feedback simulations range from $\mathcal{Q}_{\rm cold~gas, 128~x~128} = 0.78 - 0.92$. However, individual simulations often exhibit almost identical values for $\mathcal{Q}_{\rm gas}$ (e.g.\,\,the feedback runs from simulations J, UF and UP have  $\mathcal{Q}$ values higher than, but very similar to, the control runs). Furthermore, differences in $\mathcal{Q}$ of as little as 0.1 should not be taken as being significant \citep[e.g.][]{Parker14b} and based on the apparent visual structure in the feedback runs, we would expect $\mathcal{Q}$ values of less than 0.8. Given the apparent differences between the column-density images in Figures ~\ref{control_run} and ~\ref{dual_run}, this result seems counterintuitive.

\citet{Cartwright09b} and \citet{Lomax11} provide a further diagnostic of the underlying spatial distribution in relation to the $\mathcal{Q}$-parameter by plotting $\bar{m}$ against $\bar{s}$. In Fig.~\ref{cartwright_plot} we show the expected values of   $\bar{m}$ and $\bar{s}$ for distributions of $N = 10\,000$ points with various morphologies, sampling 100 versions of each morphology.  Anti-clockwise from the bottom: fractals with fractal dimension $D = 1.6$ (the black crossed circles), $D = 2.0$ (the red hashtags), $D = 2.6$ (the green dotted circles), $D = 3.0$ (the dark blue crosses); then radially smooth clusters with power law density profiles $n \propto r^{0}$ (the magenta plus symbols),  $n \propto r^{-1.0}$ (orange asterisks), $n \propto r^{-2.0}$ (blue plus symbols),  $n \propto r^{-2.5}$ (raspberry hexagrams) and $n \propto r^{-2.9}$ (purple compressed squares). $\mathcal{Q} = 0.8$ is shown by the solid line.

We plot the values of $\bar{m}$ and $\bar{s}$ for the pixel-point distributions of the gas in each of our simulations; the control runs are shown by the open symbols, and the dual feedback runs are shown by the filled symbols. Run I is shown by the triangles, Run J by the circles, Run UF by the stars, UP by the squares and Run UQ by the diamonds. The values of $\bar{m}$ and $\bar{s}$ for the control runs sit mainly around (but not on) the parameter space of the smooth, concentrated profiles with $n \propto r^{-2.0}$ (the blue plus symbols). These fake profiles typically have $\mathcal{Q}$-parameters between 0.9 -- 1.0, which is roughly consistent with the values in the simulations (0.86 -- 1.01). The runs with feedback have more scatter, and most sit roughly between the smooth, concentrated profiles with $n \propto r^{-2.0}$ and $n \propto r^{-1.0}$ (orange asterisks). The $n \propto r^{-1.0}$ profiles have $\mathcal{Q}$-parameters between 0.8 -- 0.9, again similar to the simulation values of  0.78 -- 0.91.

Despite $\mathcal{Q}$, $\bar{m}$ and $\bar{s}$ indicating a smooth, concentrated distribution, the pixel-point distributions for the gas in the simulations with feedback appear (at least to the eye) substructured.  In order to test whether this is due to a failing of the $\mathcal{Q}$-parameter in determining the structure of complex distributions, we perform two simple Monte Carlo tests in an attempt to mimic the spatial distribution in Fig.~\ref{dual_run-b}. 

\begin{figure*}
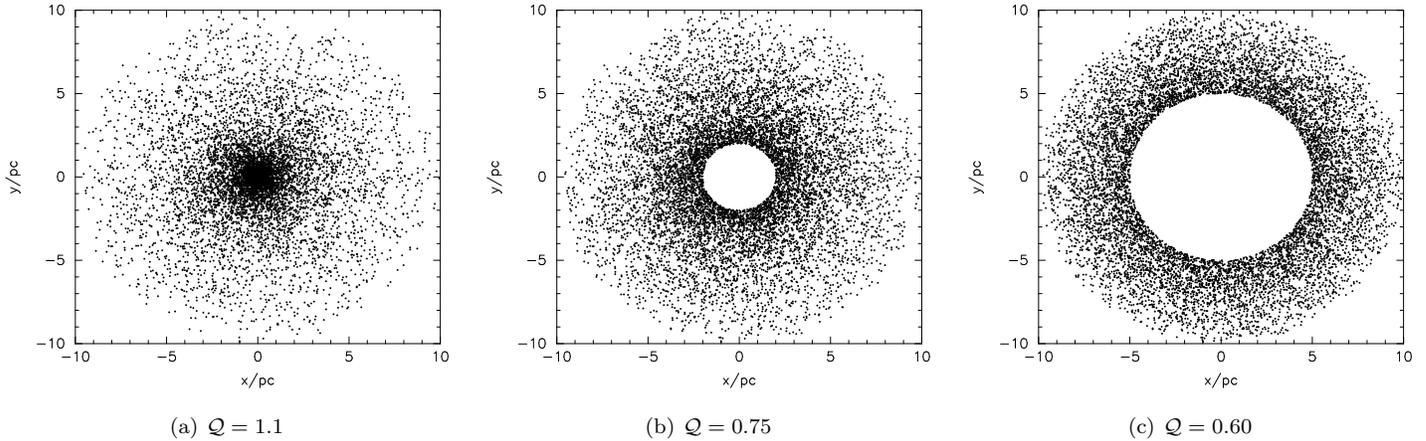

  \begin{center}
\setlength{\subfigcapskip}{10pt}
\hspace*{-1.cm}\subfigure[$\mathcal{Q} = 1.1$]{\label{ring_test-a}\rotatebox{270}{\includegraphics[scale=0.28]{2p5_N10000.ps}}} 
\hspace*{0.3cm} 
\subfigure[$\mathcal{Q} = 0.75$]{\label{ring_test-b}\rotatebox{270}{\includegraphics[scale=0.28]{2p5_N10000_p2xy.ps}}} 
\hspace*{0.3cm} 
\subfigure[$\mathcal{Q} = 0.60$]{\label{ring_test-c}\rotatebox{270}{\includegraphics[scale=0.28]{2p5_N10000_p5xy.ps}}}
\caption[bf]{The effects on the spatial distribution of a smooth, concentrated cluster with density profile $n \propto r^{-2.5}$ when removing centrally located points. In panel (a) we show the original distribution, which has $\mathcal{Q} = 1.1$. In panel (b) we have removed everything within 2\,pc of the centre and placed it at larger radii, following the underlying radial profile, and $\mathcal{Q} = 0.75$. In panel (c) we have removed everything within 5\,pc and placed them at larger radii, and $\mathcal{Q} = 0.60$.}
\label{ring_test}
  \end{center}
\end{figure*}

First, we create a smooth, concentrated distribution of 10\,000 points with radius 10\,pc and a radial density profile $n \propto r^{-2.5}$, as shown in Fig.~\ref{ring_test-a}. This has a $\mathcal{Q}$-parameter of 1.1. We then remove all points within 2\,pc of the origin, and move them to the outskirts of the distribution (thereby maintaining the same number of points and the same density profile) and $\mathcal{Q} = 0.75$ (Fig.~\ref{ring_test-b}). We then repeat this process, but remove everything within 5\,pc of the centre, and $\mathcal{Q} = 0.60$ (Fig.~\ref{ring_test-c}). This is intended to mimic in an approximate fashion the sweeping up of gas and clearing out of bubbles, which is the main visible effect of feedback. However, we keep the distribution of points smooth. Whilst the result is very similar in appearance to the gas distribution from the dual-feedback run (Fig.~\ref{dual_run-b}), it clearly has a very different underlying spatial distribution according to the $\mathcal{Q}$-parameter.

\begin{figure}
\begin{center}
\rotatebox{270}{\includegraphics[scale=0.4]{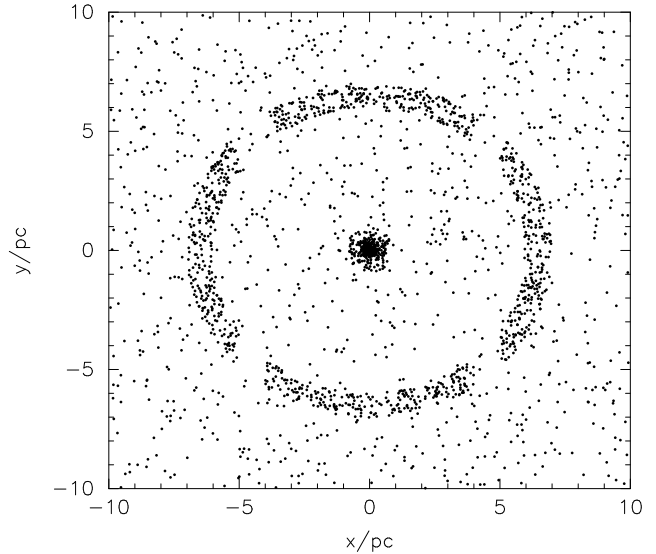}}
\end{center}
\caption[bf]{Three different spatial distributions overlaid within the same field. We show a broken ring of 2000 points within a uniform distribution of a further 1000 points. In the centre is a centrally concentrated sphere of a further 3000 points with an $n \propto r^{-2.9}$ density profile. In isolation, the broken ring has $\mathcal{Q} = 0.3$, the uniform field has $\mathcal{Q} = 0.7$ and the sphere has $\mathcal{Q} = 1.7$. The combined distributions have an overall $\mathcal{Q}$-parameter of 0.9.}
\label{Qpar_test}
\end{figure}

Secondly, we create a `broken ring' of 2000 points, which has $\mathcal{Q} = 0.3$. This is designed to represent more faithfully the broken and irregular inner walls of the feedback--driven bubbles. We then embed this in a uniform field of a further 1000 points (with  $\mathcal{Q} = 0.7$), representing the smooth and largely undisturbed background gas, and then finally, place a further 3000 points in a smooth, concentrated sphere (with an $n \propto r^{-2.9}$ density profile -- $\mathcal{Q} = 1.7$). This material represents the very dense clumps of gas found in the feedback runs. This final distribution, which incorporates all three components, is shown in Fig.~\ref{Qpar_test} and has an overall $\mathcal{Q} = 0.9$. Clearly, the concentrated clump in the centre of the distribution is dominating the overall $\mathcal{Q}$-parameter.

\begin{figure}
\begin{center}
\rotatebox{270}{\includegraphics[scale=0.4]{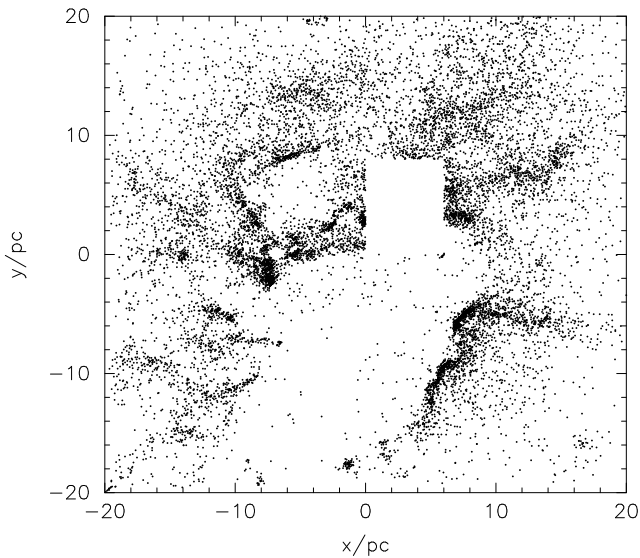}}
\end{center}
\caption[bf]{The gas pixel-point distribution for the dual-feedback Run I simulation 2.2\,Myr after feedback is enabled (Fig.~\ref{dual_run-b}), but with the pixel-points from the most dense region of gas removed. Removing this region of points reduces the $\mathcal{Q}$-parameter from 0.88 to 0.76.}
\label{hole_test}
\end{figure}

In order to test this in the SPH simulation data, we take the pixel-point distribution for the dual-feedback run shown in Fig.~\ref{dual_run-b} and remove the region of highest density centred on \{3,4\}\,pc. This distribution is shown in Fig.~\ref{hole_test}. When we remove this region, the $\mathcal{Q}$-parameter is reduced from 0.88 (indicating a smooth concentrated distribution) to 0.76, which is in the slightly substructured/uniform field regime. The remaining `background' of points, which are tracing low-density gas in the simulation, are still contributing to the $\mathcal{Q}$ value, hence the distribution still does not appear very substructured.

Our interpretation is therefore that the gas distribution in the dual-feedback runs -- whilst visually appearing to be very different from the control runs -- is actually very similar, and both are dominated by a smooth, concentrated component.

\section{Discussion}
\label{discuss}

We set out to examine the extent to which the structure of the gas and stellar distributions in a set of hydrodynamic simulations are similar using an objective quantitative criterion: the $\mathcal{Q}$--parameter. This is in contrast to many other studies where the surface density of either the stellar mass or the star formation rate are compared with the surface density of the gas, which can only offer qualitative conclusions.

If our results can be taken at face value, they suggest that the spatial distribution of stars is very different to that of the gas from which they form. In one sense, this is not surprising; the initial conditions of the simulations of \citet{Dale14} are smooth,  concentrated gas clouds, whereas stellar distributions, particularly in the early stages of star formation, are expected to be highly substructured. However, once the cloud has evolved and formed stars in a substructured distribution, we might expect the gas from which they form to also exhibit substructure. This is not the case, and the gas distribution remains dominated by a smooth concentrated component. This is particularly evident in Run I where the largest subcluster is fed by a set of smooth accretion flows which do not themselves fragment into clumps or stars.

The presence (or not) of feedback was shown in \citet{Dale12a} and \citet{Parker13a} to have a minimal effect on the spatial distribution of stars at the end of the feedback calculations, when compared with the control runs -- all the simulations from \citet{Dale12a}, \citet{Dale13} and \citet{Dale14} form stars with a substructured distribution, and other authors report similar conclusions \citep[e.g.][]{Bate12,Girichidis12,Kruijssen12a}. From visual inspection, it appears that feedback results in a very different spatial distribution for the gas (compare Figs.~\ref{control_run-b}~and~\ref{dual_run-b}). However, most of the gas remains in a dense, concentrated distribution(s) in these pixel-point maps. The difference in fact is in the relative positions of the stars and gas. In the control simulations, the densest concentrations of gas and stars are spatially coincident. In the feedback simulations, the accretion flows feeding the clusters are destroyed or deflected and the clusters and their environs are cleared of gas. Much of the gas is swept up into relatively dense concentrations, but these form stars inefficiently, so that the main effect of feedback is to spatially separate some of the stars from dense gas (this can readily be seen by comparing the positions of the sink particles to the gas in Fig.~\ref{dual_run-a}). However, in most of the runs, even in the presence of feedback many of the sink particles are still coincident with the dense gas in 2D projection, even if they not still genuinely embedded in three dimensions. This is particularly true of Run J, where large quantities of dense gas remain projected at the location of most of the stars, even though in three dimensions the main concentration of stars has been largely cleared of gas. This effect blurs out the dynamical influence of feedback when seen in projection.

One caveat is that the $\mathcal{Q}$-parameter may not be an ideal method for measuring the spatial distribution of gas, as we first need to convert a pixelated image into a distribution of points. However, \citet{Lomax11} show that the underlying spatial distribution is always recovered in synthetic datasets, and in the Monte Carlo experiment shown in Fig.~\ref{Qpar_test} we also recover the underlying spatial distribution.

\section{Conclusions}
\label{conclude}

We measure the spatial distributions of gas and stars in hydrodynamical simulations of star formation using the $\mathcal{Q}$-parameter \citep{Cartwright04,Lomax11}. We compare $\mathcal{Q}$ for simulations which form under the influence of feedback from photoionisation and stellar winds, and corresponding control run simulations. Our conclusions are as follows.

(i) The spatial distribution of the gas is different to that of the stars in all simulations. The gas has a spatially smooth, concentrated distribution ($\mathcal{Q} \sim 0.9$), whereas the stars have a substructured distribution ($\mathcal{Q} \sim 0.4 - 0.7$).

(ii) The presence of feedback clears out cavities, or bubbles in the simulations, which are visible to the eye. However, statistically these distributions are similar to the control run simulations without feedback. The reason for this is that in all simulations, a significant dense,  concentrated gas component dominates the spatial distribution.

(iii) The combination of points (i) and (ii) suggests that a direct link between the spatial distribution of young stars, and the gas from which the stars form should not necessarily be expected in observations.

These results also highlight the pitfalls in trying to measure quantitatively something previously judged by eye. The column density images shown in Figures \ref{control_run-a} and \ref{dual_run-a} and the corresponding pixel-point distributions could scarcely look more different, yet the $\mathcal{Q}$-parameter analysis reports that they are statistically indistinguishable. The root cause of the similarity is that they are both dominated by the same kinds of structures, the visual difference being due to how those structures are arranged.

There is growing evidence in the literature that suggests star formation is hierarchical, with no preferred spatial scale \citep{Efremov95,Elmegreen06b,Bastian07,Kruijssen12b}. Because of this, it is tempting to link different regimes together, such as the spatial distribution of gas in the ISM with the spatial distribution of stars in clusters \citep[e.g.][]{Gouliermis14,Hetem15}. Our results suggest that -- whilst star formation in numerical simulations also produces a hierarchical spatial distribution of the stars -- the distributions of gas and stars may be linked in a highly non-trivial way.

\section*{Acknowledgments}

We thank the referee, Olly Lomax, for helpful suggestions which improved the original manuscript. RJP acknowledges support from the Royal Astronomical Society in the form of a research fellowship, and from the European Science Foundation (ESF) within the framework of the ESF `Gaia Research for European Astronomy Training' exchange visit programme (exchange grant 4994). This research was also supported by the DFG cluster of excellence `Origin and Structure of the Universe' (JED). 

\bibliography{general_ref}

\label{lastpage}

\end{document}